\documentclass[aps,prb,twocolumn,superscriptaddress,10pt]{revtex4-2}
\usepackage{graphicx}
\usepackage{amsmath}
\usepackage{comment}
\usepackage{color}
\usepackage{soul}
\usepackage[T1]{fontenc} 

\begin{document}

\title{Relation of Continuous Chirality Measure to Spin and Orbital Polarization, and Chiroptical Properties in Solids}

\author{Andrew Grieder}
\affiliation{Department of Materials Science and Engineering, University of Wisconsin-Madison, WI, 53706, USA}
\author{Shihao Tu}
\affiliation{Department of Materials Science and Engineering, University of Wisconsin-Madison, WI, 53706, USA}
\author{Yuan Ping}
\email{yping3@wisc.edu}
\affiliation{Department of Materials Science and Engineering, University of Wisconsin-Madison, WI, 53706, USA}
\affiliation{Department of Physics, University of Wisconsin-Madison, WI, 53706, USA}
\affiliation{Department of Chemistry, University of Wisconsin-Madison, WI, 53706, USA}


\begin{abstract}
    
Chirality introduces intriguing topological, electronic, and spin-optronic properties to molecules and solids. In this work, we provide a quantitative metric for the degree of chirality in solids, independent of the type of system and the dimensionality, through the continuous chirality measure (CCM). We quantitatively analyze the correlation between CCM and spin and orbital angular momentum (OAM) polarization, as well as circular dichroism (CD) and the circular photogalvanic effect (CPGE). By internal spin-orbit field analysis, we demonstrate a distinct character (proportionality among Rashba, Deresselhaus, and Weyl contributions) and chirality dependence among different chiral solids. Furthermore, unlike CD, we found that absorption dissymmetry factor $g_{CD}$ could remain unchanged as a function of chirality and show anisotropic dependence on CCM. In addition, we show that the relation between CCM and CPGE is rather complex. At low excitation energy close to the bandgap transition, the CCM continuously tunes the total SOC, and therefore, the CPGE response. However, at high excitation energy, CPGE includes more than just band edge transitions, which complicates the relation of chirality and CPGE due to changes in the optical dipole strength and electron-hole group velocity difference. Ultimately, this causes CPGE to be only correlated with chirality at excitation energies close to the band edge. 
At the end, we discussed strategies of manipulating chiral-optical properties through chirality transfer at interfaces or applying strain. 
The insights developed in this work will inspire the design of materials for future spintronics and orbitronics, as well as spin-optronics applications. 

\end{abstract}

\maketitle
\section{Introduction}

Structural chirality lacks spatial inversion and mirror symmetries, resulting in distinct electronic and optical responses~\cite{siegel1998homochiral}. This inherent asymmetry plays a crucial role in optical orientation, where chiral materials interact differently with left- and right-circularly polarized light. Such interactions give rise to circular dichroism (CD)~\cite{kelly2005study} and circularly polarized luminescence (CPL)~\cite{mun2020electromagnetic,sang2020circularly}. The ability to control light-matter interactions through chirality enhances spin-dependent optical functionalities in modern optoelectronic applications. Besides optical activity, chirality has important implications in spin transport; for example large spin polarization can be generated by running charges in nonmagnetic low-spin-orbit chiral materials, known as the chirality-induced spin selectivity (CISS) effect~\cite{naaman2012chiral,naaman2019chiral}. Such effect implies spontaneous time-reversal symmetry broken in absence of permanent magnetization. Since its discovery in 2012~\cite{naaman2012chiral}, researchers have proposed different theories to explain its mechanism, but no clear agreement has been reached~\cite{calavalle2022gate,roy2022long,tenzin2023collinear}. Additionally, chirality plays a key role in orbitronics, an emerging field that utilizes electron orbital angular momentum (OAM)~\cite{liu2021chirality,kim2023optoelectronic,yang2023monopole,brinkman2024chirality} as an information carrier. The broken inversion symmetry in chiral materials enables 
coupling between OAM and spin angular momentum (SAM)~\cite{hirayama2015weyl,sakano2020radial,gatti2020radial,yang2025chirality}, leading to 
chiral orbital textures and potentially large orbital Hall conductivity (OHC)~\cite{brinkman2024chirality,go2018intrinsic}. By integrating these effects, chiral materials serve as a versatile platform for next-generation optoelectronic, spintronic, and orbitronic applications, enabling advances in quantum information science and low-power electronics~\cite{yang2021chiral,forment2022chiral}.

Quantifying chirality is critical to understanding and controlling its effects on electronic and optical properties. The Continuous Chirality Measure (CCM) offers a reliable way to do so. Unlike binary classifications of chirality, CCM provides a continuous scale to evaluate the degree of chirality in a given structure. For example, Hausdorff distances and CCM have been applied to crystal structures to distinguish chiral space groups and analyze their influence on band structures and Berry curvatures~\cite{fecher2022chirality}. In another approach, graph-theoretical chirality (GTC)~\cite{cha2024graph}, based on torsion in differential geometry, has been proposed to describe chirality across different length scales, offering direct correlations with optical activity and chiral meta-materials. Beyond structural characterization, CCM has emerged as a crucial parameter in describing the electronic properties of chiral materials, in particular to their spin and orbital textures, as we will show in this work. In systems with the presence of CISS, a larger CCM often correlates with stronger spin-polarization (or spin-splitting in $\textbf{k}$ space), suggesting a direct impact on spin-selective transport. Similarly, in chiral crystals, OAM textures emerge and their properties can be quantitatively related to CCM, as shown in this work. Thus, CCM not only measures chirality geometrically but also predicts emergent quantum properties~\cite{Chang2018-jj}.

CCM plays a crucial role in linking structural chirality to chiroptical properties such as CD and the Circular Photogalvanic Effect (CPGE). Since CD reflects differential absorption of left- and right-circularly polarized light, higher CCM intuitively correlates with a stronger CD signal. This is because a higher CCM may correspond to more pronounced chiral electronic states or electronic transition properties, thereby enhancing the optical activity. Experimentally, the absorption dissymmetry factor ($g_{CD}$) is often used to characterize the degree of chirality, which normalizes the CD by the optical absorption. However, the correlation between $g_{CD}$ and the degree of chirality remains to be validated, and is examined in this work. Recent studies demonstrate that tuning chiral molecule-surface interactions can realize chirality transfer to nonchiral surfaces and enhances CD,
reinforcing the connection between CCM and chiroptical properties~\cite{Haque2024-yw, song2024enhancing,apergi2023calculating}. 

Besides CD, CPGE is another common optical characterization technique for chiral materials. With circularly-polarized light excitation, DC photocurrent is generated in chiral solids, and governed by characteristic symmetry relations~\cite{le2021topology}. It also  
has been used to probe the topological charge of Weyl points in chiral materials~\cite{de2017quantized,le2020ab}. 
Furthermore, CPGE offers a unique probe of chirality-sensitive charge dynamics, as its response is inherently linked to broken crystal symmetry and spin-orbit coupling (SOC) effects~\cite{yu2015temperature,zhu2024electrically}. Since chirality breaks inversion symmetry, these SOC-induced band splittings can lead to spin-polarized photocurrents under circularly-polarized light. This makes CPGE a promising tool for characterizing chiral materials, providing complementary insights beyond traditional chiroptical measurements like CD, particularly in chiral semiconductors and 2D hybrid organic-inorganic perovskites (2D HOIP)~\cite{zhu2024electrically}. Despite its importance, the relation between CPGE and the degree of chirality remains elusive and requires in-depth investigations. 

Previous works have investigated the correlation between structural properties and SOC spin-splitting as well as CD in 2D-HOIPs. These studies have focused on structural parameters unique to halide perovskites~\cite{jana2021structural,chakraborty2024design, fortino2024role, pols2024temperature, apergi2023calculating}.  For example, one measure is the $\Delta \beta_{in}$, which is the difference in M-X-M bond angles projected to the plane of the 2D inorganic sublattice~\cite{jana2021structural}. They find a correlation between $\Delta \beta_{in}$ and the Rashba parameters, but a later study found $\Delta \beta_{in}$ is not suitable for \textit{Aba2} and \textit{Pbcn} space-groups, where $\Delta \beta_{in}=0$ despite large spin splittings~\cite{chakraborty2024design}. In another study, CD was computed from the tight-binding theory and did not correlate with $\Delta \beta_{in}$ either~\cite{apergi2023calculating}. Instead they correlated CD with the relative distances of neighboring axial halide atoms with the center atom of the octahedral cage. Other studies have used \textit{ab initio} molecular dynamics  (AIMD) to investigate temperature effects on chiral perovskites~\cite{pols2024temperature,fortino2024role}. They found a reduction in chirality as temperature increases~\cite{pols2024temperature}. Additionally, tin-based perovskites have a larger distortion due to weaker bond affinity ~\cite{fortino2024role}. While past works focus on specific structural properties of 2D-HOIPs, here we focus on using a universal metric for chiral molecules and solids, i.e. continuous-chirality measure, to investigate its relation to spin and chiral-optical properties across different classes of materials.  
To this end we choose two disparate prototypical chiral systems: one 2D HOIP (2D-NPB) and one inorganic single-element solid, Se. Instead of including a large number of systems, we aim for an in-depth analysis of these two materials, which have distinct symmetry, chemical bonding, potentially representative of a larger class of materials. Additionally they have comprehensive experimental studies for theoretical validation. 
In contrast to previous studies, we quantify the relation of the degree of chirality with a more comprehensive set of observables, e.g. including CPGE and OAM for the first time.

In addition to inversion symmetry breaking, perovskites are known to have dynamic disorder that induces an effect such as dynamical Rashba ~\cite{li2024spin, abramovitch_thermal_2021}.  Specifically, in the centrosymmetric perovskite MAPbBr$_3$, the upper limit of dynamical Rashba splitting was found to be 1.35 eV/\AA ~\cite{li2024spin}. This indicates the significance of this effect in perovskites. While this is indeed an area of research that warrants further attention, it is outside the scope of this work. Instead, we focus on the effects of chirality alone. Additionally, we find excellent agreement with experimental CD for 2D-NPB (except for the missing excitonic peak below the bandgap transitions); see Figure~\ref{fig:Exp-Benchmark}d. Therefore, we believe that the contribution from chirality-induced symmetry breaking significantly outweighs the effects from dynamical disorder.

In this work we present a quantitative assessment of relation between the degree of chirality (CCM) and a comprehensive set of spin-optotronic properties (CD, $g_{CD}$, CPGE, OAM, and spin) across chiral materials. The remainder of this paper is organized as follows. In Section 2.1, we introduce the definition of CCM for solid-state crystals and two-dimensional hybrid perovskites, establishing a quantitative framework for measuring the degree of chirality and exploring its correlation with spin and orbital textures in these systems. In Section 2.2, we investigate the influence of chirality on optical responses, specifically CD and absorption dissymmetry factor ($g_{CD}$). We discuss the anisotropic dependence of $g_{CD}$ on chirality and further examine the role of strain in modulating CD. In Section 2.3, we analyze the relationship between chirality and CPGE, highlighting the complex factors determining CPGE beyond SOC spin-splitting. In Section 2.4, we explore the transfer of chirality across organic-inorganic interfaces in hybrid perovskites, emphasizing how this process modifies both electronic and optical properties, particularly in enhancing CD and CPGE. Finally, in Section 3, we summarize our findings and propose future directions for leveraging chirality in the design of next-generation optoelectronic, spintronic and orbitronic materials.

\section{Result and Discussion}
\subsection{Continuous Chirality Measure}

The CCM has been commonly used to quantify the chirality in molecules and was recently adapted to describe periodic crystals \cite{fecher2022chirality}. CCM is defined by the mean squared distance between atomic positions in the chiral structure and their closest equivalent positions in a symmetric reference structure. This approach ensures that the measure effectively captures the degree of chirality. The unnormalized chirality measure is defined as:
\begin{equation}
\tilde{S}^2(G) = \sum_{i=1}^{n} \left\| p_i - p_i^{\text{sym}} \right\|^2,
\end{equation}

where \( p_i \) are the atomic positions in the target chiral structure and \( p_i^{\text{sym}} \) are the closest equivalent positions in an achiral reference space group ($G$) such that $\tilde{S}^2(G)$ is minimized.

To compare different structures, the measure is normalized as:
\begin{equation}
S^2(G) = \frac{\tilde{S}^2(G)}{N} \times 100,
\end{equation}

where \( N \) is a normalization factor that accounts for the difference between the achiral and chiral structures:
\begin{equation}
N = \sum_{i=1}^{n} \left\| p_i^{\text{chiral}} - p_i^{\text{sym}} \right\|^2.
\end{equation}
The positions of the chiral reference structure ($p^{\text{chiral}}$) are determined such that $N$ is maximized. Normalization allows \( S^2(G) \) to vary from 0 (no chirality) to 100\% (fully chiral). A prototypical system for studying the change in CCM is trigonal selenium (Se).

\begin{figure}
    \centering\includegraphics[width=\linewidth]{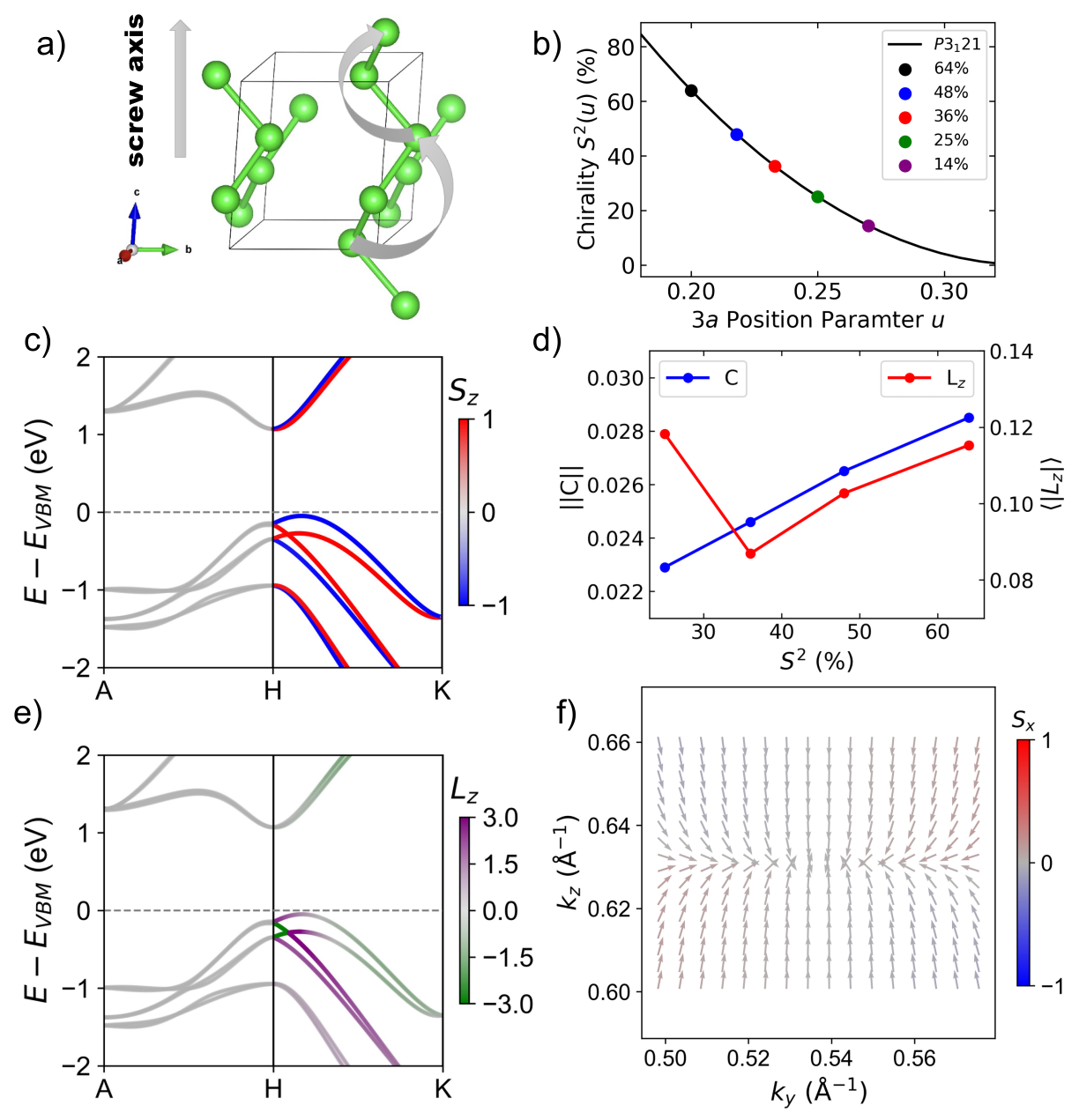}
    \caption{Spin and orbital texture of chiral Se as a function of chirality. (a) Chiral Se Structure with $P3_121$ symmetry.(b)  Normalized Continuous Chirality Measure ($S^2$) as a function of the $3a$ Wycoff positions for $P3_121$ symmetry. (c) and (e) are the Se band structure with bands colored by $\langle S_z \rangle$ and $\langle L_z \rangle$ respectively. (d) Fermi-Dirac average of $\langle |L_{z}| \rangle$ and spin splitting C as a function of chirality. (f) Se Spin texture centered at the H high-symmetry point in the $k_y$-$k_z$ plane parallel to the screw axis z.}
    \label{fig:CCM-and-Bands}
\end{figure}

 Trigonal Se crystallizes in a chiral structure with space group \( P 3_1 2_1 \) (No.152) or its enantiomorphic counterpart \( P 3_2 2_1 \) (No.154). The primitive unit cell consists of a helical atomic chain arranged along the 3-fold screw axis, forming a right-handed or left-handed screw symmetry, as shown in Figure~\ref{fig:CCM-and-Bands}a. The Wyckoff positions of the atomic sites define the chirality of the structure: in \( P 3_1 2_1 \), the atoms are located at \( (u, 0, \frac{1}{3}\)), \( ( 0,u, \frac{2}{3}\)), and \( (-u, -u, 0\)), whereas in \( P 3_2 2_1 \), they shift to \( (u, 0, \frac{2}{3}) \), \( ( 0,u, \frac{1}{3}\)), and \( (-u, -u, 0\)). With fixed lattice constants at experimental values, we evaluate \(u\) values of 0.200, 0.218 (experimental), 0.233, 0.250, and 0.270, with the corresponding chirality measures \( S^2 \) of 64\%, 48\%, 36\%, 25\% and 14\%, respectively. The structural evolution along these paths is shown in Figure~\ref{fig:CCM-and-Bands}b, exhibiting a trend of decreasing chirality. Such continuous tuning of the degree of chirality in Se can be realized experimentally by applying strain. 

A second prototypical chiral system we investigate is two-dimensional (2D) hybrid organic-inorganic perovskites (HOIPs). 
Incorporation of chiral organic cations into 2D HOIPs results in a transfer of structural chirality to inorganic lattice, by inducing chiral distortion~\cite{jana2020organic}. For example, R-(+)- and S-(-)-1-(1-naphthyl)ethylammonium (R/S-NEA) cations have been incorporated into layered lead bromide perovskites, forming (R/S-NEA)\textsubscript{2}PbBr\textsubscript{4} structures (R/S-NPB). These chiral spacers induce symmetry-breaking distortions in the inorganic framework through asymmetric hydrogen bonding, an effect not observed in the racemic system. This results in the P2\textsubscript{1} chiral symmetry, leading to CD and Rashba-Dresselhaus spin splitting. To investigate the role of chirality transfer we generate configurations through a structural interpolation between the centrosymmetric structure and one with full chiral-distortion. To simplify structural interpolation, we replace the chiral molecules with Cs\textsuperscript{+}, which also removes any electronic effects resulting from the molecules (we note no Cs related states close to band edges). The initial reference structure (CCM = 0\%) is an undistorted perovskite lattice and corresponds to the inorganic sublattice of racemic-NPB. The final structure (CCM = 100\%) is the experimentally observed distorted state, where chiral organic cations cause a clear helical deformation in the inorganic layers, breaking mirror and inversion symmetries. To distinguish the hybrid structure with molecules and the interpolated structure with Cs, we adopt the notation of 2D-NPB and 2D-Cs-PB, respectively. To obtain structures with intermediate degree of chirality, we use a simple linear interpolation method based on atomic positions and lattice parameters. This method gradually changes the reference structure into the chiral distorted structure by smoothly adjusting atomic positions, bond angles, and torsional distortions along a defined path. Specifically, the interpolation follows:

\begin{equation}
R_j(\lambda) = R_j^{(i)} + \lambda (R_j^{(f)} - R_j^{(i)})
\end{equation}
where \( R_j(\lambda)\) is the atomic position of the \(j\)-th atom at an interpolation step (ranging from 0 to 1), \(R_j^{(i)}\) is the atomic position in the initial centrosymmetric reference structure, and \(R_j^{(f)}\) is the final experimental chiral structure.

\subsection{Spin and Orbital Texture}

\begin{figure}
    \centering
    \includegraphics[scale=0.35]{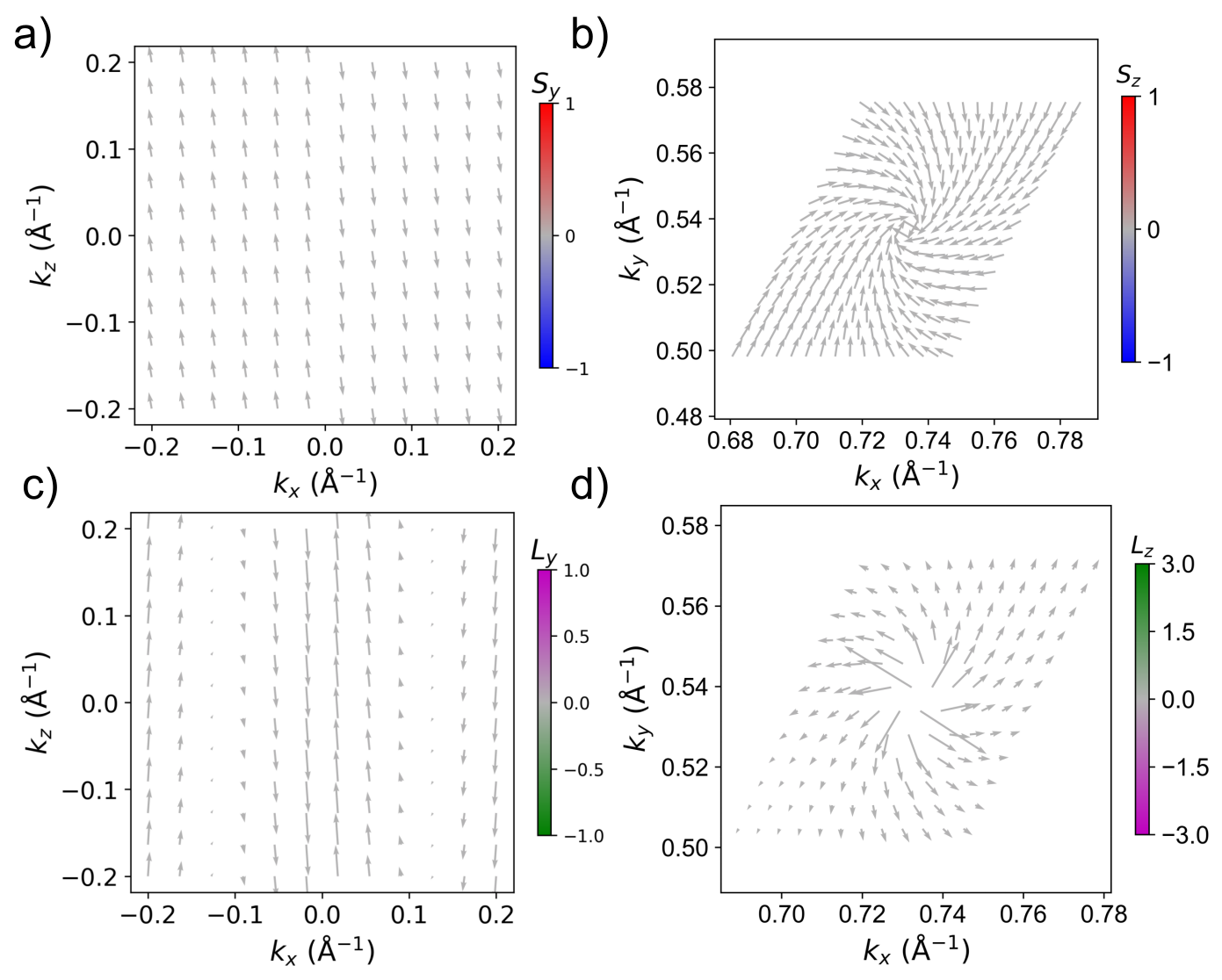}
    \caption{Spin and orbital texture of chiral solids. (a) Spin texture in the $k_x$-$k_z$ plane centered at $\Gamma$ high-symmetry point of the CBM in 2D-NPB  (z is the layer-spacing direction) and  (b) spin texture in the $k_x$-$k_y$ plane centered at H high-symmetry point of the CBM in Se, normal to the screw axis z. Orbital texture in the same planes for (c) 2D-NPB and  (d) Se. Color bars indicate the out-of-plane component of spin/orbital polarization.}
    \label{fig:Spin-Orb-Texture-PVK-Se}
\end{figure}

One key property of interest in chiral structures is the presence of orbital and spin polarization and textures, thus we systematically investigate how these properties change with CCM. We find both orbital and spin textures are strongly correlated with the degree of chirality. The Se crystal structure with SG \textit{P}3$_1$21 possesses 3-fold screw rotation \texttt{\{}$C_{3z}$~$\mid$~0, 0, $\frac{1}{3}$\texttt{\}}, and 2-fold rotations \texttt{\{}$C_{2y}$~$\mid$~0, 0, $\frac{1}{3}$\texttt{\}} and \texttt{\{}$C_{2, 110}$~$\mid$~0, 0, 0\texttt{\}} as the symmetry operators. In the following, the term "screw axis" specifically refers to the 3-fold screw rotation along the z-direction. These symmetries impose constraints on the SAM and OAM in momentum space, causing certain directional components to vanish along specific high-symmetry paths. Specifically, along the $\Gamma$--A and K--H lines, $\mathrm{S}_{x,y}(\mathbf{k})$ and $\mathrm{L}_{x,y}(\mathbf{k})$ vanish due to symmetry constraints, while $\mathrm{S}_z(\mathbf{k})$ and $\mathrm{L}_z(\mathbf{k})$ are nonzero with z chosen to be along the screw axis. In contrast, along the $\Gamma$--K--M and A--H--L lines, $\mathrm{S}_{x,y}(\mathbf{k})$ and $\mathrm{L}_{x,y}(\mathbf{k})$ are nonzero, whereas $\mathrm{S}_z(\mathbf{k})$ and $\mathrm{L}_z(\mathbf{k})$ vanish (as shown in Figure~\ref{fig:CCM-and-Bands} and SI Figure S1). We mainly focus on the SAM and OAM distribution at the conduction band minimum (CBM) of the H point, as it is a crucial Weyl point~\cite{hirayama2015weyl}. The SAM and OAM textures along the high-symmetry lines passing through the H point are shown in the Figure~\ref{fig:CCM-and-Bands}c and e. The results are consistent with the analysis based on symmetry constraints. The complete SAM and OAM-resolved band structures and textures of Se can be found in SI Figures S1 and S4. 
In general, we found a good correlation between spin and orbital textures in k-space in chiral systems. For a periodic system, the OAM is computed based on the Berry-phase formalism~\cite{Xu2024-lw}:
\begin{equation}
      \mathbf{L}_{\mathbf{k},nm}=i \left\langle \frac{\partial u_{\mathbf{k},n}}{\partial \mathbf{k}} 
      \left|\widehat{H}-\frac{\epsilon_{\mathbf{k},n}+\epsilon_{\mathbf{k},m}}{2} \right|
      \frac{\partial u_{\mathbf{k},m}}{\partial \mathbf{k}}\right\rangle,
\end{equation}

where $\widehat{H}$ is the Hamiltonian operator, $u_{\mathbf{k},n}$ is the periodic part of the Bl\"och wavefunction, $n$ and $m$ are band indices, and $\epsilon$ is the corresponding eigen-energy. Details of the implementation can be found in Refs.~\cite{multunas2023circular,Xu2024-lw}. The computed OAM has good agreement with experimental CD-ARPES measurements, see computational methods section.
Correlation of spin and orbital textures for Se and 2D-NPB are highlighted in Figure~\ref{fig:Spin-Orb-Texture-PVK-Se}. A persistent spin helix (PSH) like spin texture can be seen for both the spin and orbital textures in 2D-NPB, Figure~\ref{fig:Spin-Orb-Texture-PVK-Se}a and c. We later will validate the presence of PSH through fitting with an effective SOC Hamiltonian.
Interestingly, the orbital texture flips sign around below 0.12 \AA$^{-1}$, becoming anti-parallel to the spin texture. In contrast, Se has a more radial spin texture in the plane normal to the screw axis as discussed below and we see the similar pattern in the orbital texture, Figure~\ref{fig:Spin-Orb-Texture-PVK-Se}b and d. In Se, parallel to the screw axis, the spin texture is not radial, but with a parallel spin texture, shown in Figure~\ref{fig:CCM-and-Bands}f. Normal to the screw axis, there is no out of plane component to the spin or orbital textures for both Se and the 2D-Cs-PB system, indicated by the absence of any color in Figure~\ref{fig:Spin-Orb-Texture-PVK-Se}.

\begin{figure}
    \centering
    \includegraphics[scale=0.48]{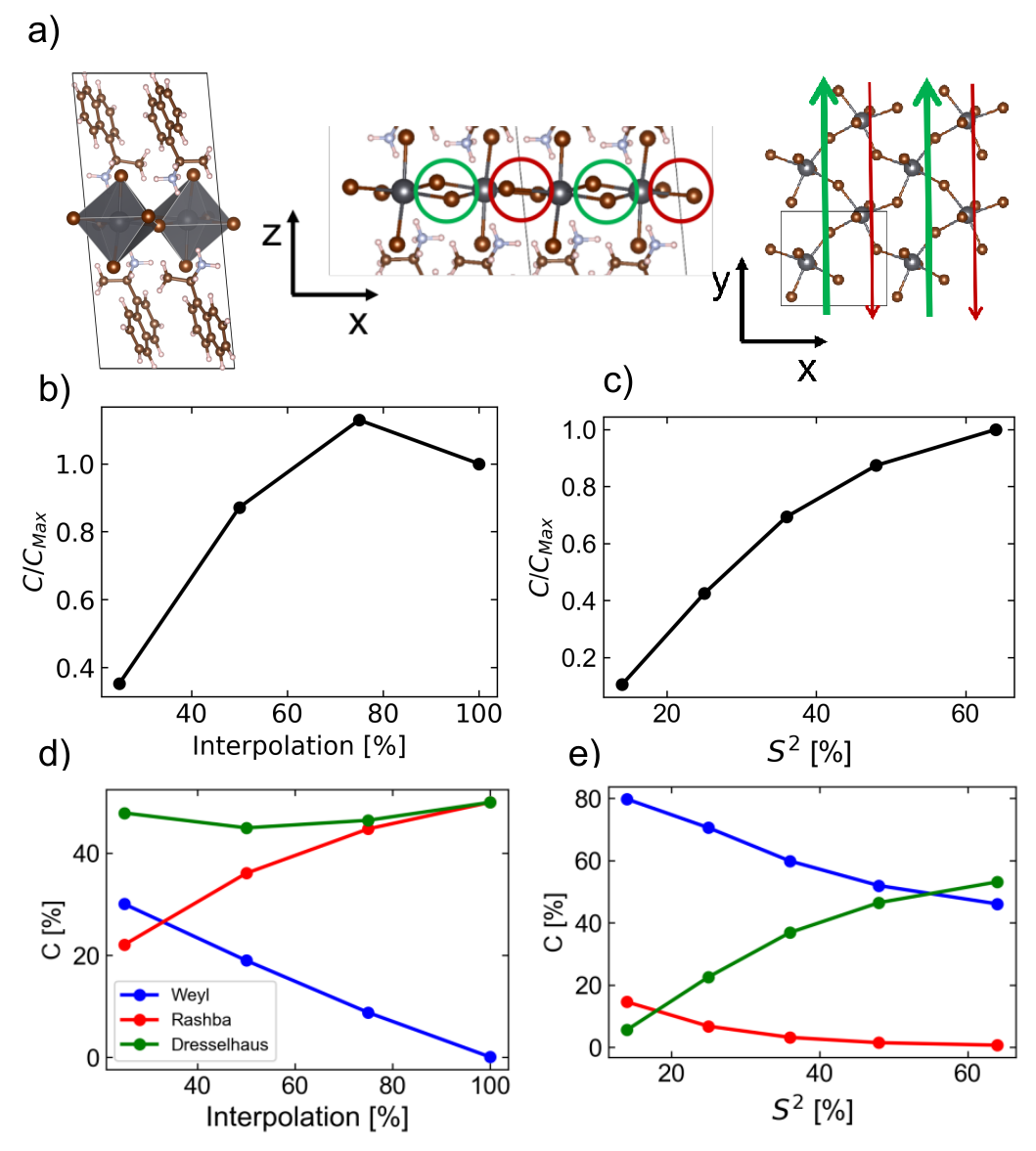}
    \caption{Spin texture analysis for Weyl, Rashba, and Dresselhaus contributions as a function of chirality for two systems. (a) 2D-NPB with major and minor $P2_1$ screw axis highlighted in green and red, respectively. Normalized magnitude of fitting matrix $C$ for (b) 2D-Cs-PB and (c) Se. Contributions from Weyl, Rashba, and Dresselhaus spin types to the spin splitting (d) 2D-Cs-PB and (e) Se.}
    \label{fig:Spin-texture-C-fitting}
\end{figure}

In order to quantify spin texture we employ an effective Zeeman Hamiltonian as in Ref.~\cite{li2024spin}:

    \begin{equation}
        H^{\text{in}}=\sigma^T \cdot \mathbf{B}^{\text{in}}(\textbf{k})
    \end{equation}
    \begin{equation}
        \mathbf{B}^{\text{in}}=\frac{2\Delta_{\textbf{k}}\mathbf{S_\textbf{k}}}{g_0\mu_B}=C\cdot \boldsymbol{\textbf{k}}
        \label{Bin-CK}
    \end{equation}  
    \begin{equation}
        C=C_s + C_v + C_t , \label{zeemanfit}
    \end{equation}
    
where $\sigma$ are the Pauli matrices, $\mu_B$ is the Bohr magneton, $g_0$ is the free electron $g$ factor, $\Delta_\textbf{k}$ is the spin splitting energy of the bands, $\mathbf{S}_\textbf{k}$ is the spin angular momentum, $\mathbf{B}^{\text{in}}(\textbf{k})$ is the internal effective magnetic field originating from SOC.  
With Eq.~\ref{Bin-CK}, the coefficient tensor $C$ can be extracted by fitting least squares to the calculations of first principles of $\mathbf{S}_\textbf{k}$ and $\Delta_\textbf{k}$. The $C$ tensor can then be decomposed, enabling the quantification of different contributions to the spin texture (assuming linear in k close to the band minimum). The Weyl contribution ($C_s$) is the trace, the Rashba contributions ($C_v$)  is the traceless antisymmetric part, and the Dresselhaus contributions ($C_t$) is the symmetric part; see SI section II for more details~\cite{li2024spin}. For both systems tested in Figures~\ref{fig:Spin-texture-C-fitting}b and c, the magnitude of $C$ (or the magnitude of $\mathbf{B}^{\text{in}}$) increases with chirality. This directly reflects the relation between symmetry-broken and strength of spin-orbit field. 
Although overall the magnitude of $C$ increases with increasing chirality, consistent with increased symmetry-broken, the relative contributions vary between the 2D-Cs-PB and chiral Se due to their different crystal symmetry. 

In chiral Se,  the spin texture was found to consist primarily of the Weyl and Dresselhaus terms, as shown in Figure~\ref{fig:Spin-texture-C-fitting}e. At low chirality, the Weyl contribution is dominant, and the Dresselhaus component increases with increasing chirality. The trend is consistent with a qualitative analysis of the spin texture normal to the screw axis (see Figure~\ref{fig:Spin-Orb-Texture-PVK-Se}b,  which shows a predominantly radial pattern with a certain degree of radial-tangential mixing). According to the classification summarized by Zunger \textit{et al.}~\cite{mera2021different}, such a texture indicates a hybrid spin polarization character that originates from a mixture of Weyl and Dresselhaus terms. However, the increasing Dresselhaus contribution may be missed from a qualitative analysis. Similarly to the spin texture, the orbital texture also exhibits the radial-tangential mixing for all degree of chirality; see SI Figure S4.

In the 2D-Cs-PB structure at low chirality, all three spin texture types are present in comparable magnitudes (Figure~\ref{fig:Spin-texture-C-fitting}d). Similarly to Se, the Weyl contribution decreases as the chirality increases. However, in contrast to Se, the Rashba contribution increases with increasing chirality. 
At the highest chirality, it shows an equal contribution of Rashba and Dresselhaus and satisfies the persistent spin helix (PSH)~\cite{Bernevig2006, li2024spin} Hamiltonian $\mathbf{H}_{\text{PSH}}=\alpha'k_x\sigma_z$,
resulting in a nearly perfect PSH texture as shown in Figure~\ref{fig:Spin-texture-C-fitting}d.  Similar to Se, the orbital texture closely resembles the spin texture as shown in SI Figure S3. Because the degree of chirality in the halide perovskite is dictated by the interaction with chiral molecules, one can experimentally tune the spin and orbital textures by engineering chiral molecules.

\subsection{Effect of Chirality on CD}

Optical activity such as Circular Dichroism (CD) is a direct probe for chirality.  
We will discuss the quantitative relation between CD and the degree of chirality in the following. 
The CD response was computed based on the Berry phase formalism as detailed in the previous study~\cite{multunas2023circular}, 
including both magnetic dipole (MD) and electric quadrupole (EQ) contributions: 

\begin{equation}
\begin{aligned}
\Delta \alpha_{\mu\nu} =& \\&
\frac{4 \pi^2 e^2}{m_e^2 \omega^2} 
\int \frac{g_s \, d\mathbf{k}}{(2\pi)^3} 
\sum_{n',n}
\delta \!\left( \varepsilon_{\mathbf{k}n'} - \varepsilon_{\mathbf{k}n} - \hbar \omega \right) 
\left( f_{\mathbf{k}n} - f_{\mathbf{k}n'} \right)
\\&\times \mathrm{Re}\,\mathrm{Sym} \Bigg[
 \delta_{\mu\nu} \, p_{\rho,n'n}^{\mathbf{k}\,*} 
 \left( L_{\rho,n'n}^{\mathbf{k}} + 2 S_{\rho,n'n}^{\mathbf{k}} \right) 
 \\&- p_{\mu,n'n}^{\mathbf{k}\,*} 
 \left( L_{\nu,n'n}^{\mathbf{k}} + 2 S_{\nu,n'n}^{\mathbf{k}} \right) 
 + \epsilon_{\mu\nu\rho} \, p_{\sigma,n'n}^{\mathbf{k}\,*} 
 Q_{\rho\sigma,n'n}^{\mathbf{k}}
\Bigg]~,
\end{aligned}
\label{CD_equation}
\end{equation}

where Latin indices denote Cartesian directions; $n$, $\mathbf{k}$, and $f$  are the band index, the Bloch wave vector in the first Brillouin zone, and the Fermi-Dirac distribution respectively. $\omega$  is the photon frequency, $g_s$ is the spin degeneracy, and $p$ are the elements of the momentum matrix. Additionally, $L$ are magnetic dipole matrix elements, $Q$ electric quadrupole matrix elements, and $S$ spin matrix elements.  
Here $\mathrm{Sym}[F_{\mu\nu}]=\frac{F_{\mu\nu} + F_{\nu\mu}}{2}$, for more details see Ref.~\cite{multunas2023circular}. For convenience of discussions, we separate the MD, EQ, and spin contributions to CD as follows. First, the MD contribution to CD is defined as:

\begin{equation}
\begin{aligned}
&CD_{MD}[\mu\nu] = \\&
\frac{4 \pi^2 e^2}{m_e^2 \omega^2} 
\int \frac{g_s \, d\mathbf{k}}{(2\pi)^3} 
\sum_{n',n}
\delta \!\left( \varepsilon_{\mathbf{k}n'} - \varepsilon_{\mathbf{k}n} - \hbar \omega \right) 
\left( f_{\mathbf{k}n} - f_{\mathbf{k}n'} \right)
\\& \times \mathrm{Re}\,\mathrm{Sym} \Bigg[
 \delta_{\mu\nu} \, p_{\rho,n'n}^{\mathbf{k}\,*} 
 \left( L_{\rho,n'n}^{\mathbf{k}} + 2 S_{\rho,n'n}^{\mathbf{k}} \right) 
 \\&- p_{\mu,n'n}^{\mathbf{k}\,*} 
 \left( L_{\nu,n'n}^{\mathbf{k}} + 2 S_{\nu,n'n}^{\mathbf{k}} \right) 
\Bigg]~,
\end{aligned}
\label{CD_MD_equation}
\end{equation}

the EQ contribution to CD is defined as:

\begin{equation}
\begin{aligned}
&CD_{EQ}[\mu\nu] = \\&
\frac{4 \pi^2 e^2}{m_e^2 \omega^2} 
\int \frac{g_s \, d\mathbf{k}}{(2\pi)^3} 
\sum_{n',n}
\delta \!\left( \varepsilon_{\mathbf{k}n'} - \varepsilon_{\mathbf{k}n} - \hbar \omega \right) 
\left( f_{\mathbf{k}n} - f_{\mathbf{k}n'} \right)
\,\\& \times \mathrm{Re}\,\mathrm{Sym} \Bigg[\epsilon_{\mu\nu\rho} \, p_{\sigma,n'n}^{\mathbf{k}\,*} 
 Q_{\rho\sigma,n'n}^{\mathbf{k}}
\Bigg]~,
\end{aligned}
\label{CD_EQ_equation}
\end{equation}

and the spin contribution to CD is defined as:

\begin{equation}
\begin{aligned}
&CD_{S}[\mu\nu] = \\&
\frac{4 \pi^2 e^2}{m_e^2 \omega^2} 
\int \frac{g_s \, d\mathbf{k}}{(2\pi)^3} 
\sum_{n',n}
\delta \!\left( \varepsilon_{\mathbf{k}n'} - \varepsilon_{\mathbf{k}n} - \hbar \omega \right) 
\left( f_{\mathbf{k}n} - f_{\mathbf{k}n'} \right) \times
\,\\&\mathrm{Re}\,\mathrm{Sym} \Bigg[
 \delta_{\mu\nu} \, p_{\rho,n'n}^{\mathbf{k}\,*} 
 \left(2 S_{\rho,n'n}^{\mathbf{k}} \right) 
 - p_{\mu,n'n}^{\mathbf{k}\,*} 
 \left(2 S_{\nu,n'n}^{\mathbf{k}} \right) 
\Bigg]~.
\end{aligned}
\label{CD_MD_equation2}
\end{equation}

\begin{figure}[!hb]
    \centering
    \includegraphics[scale=0.42]{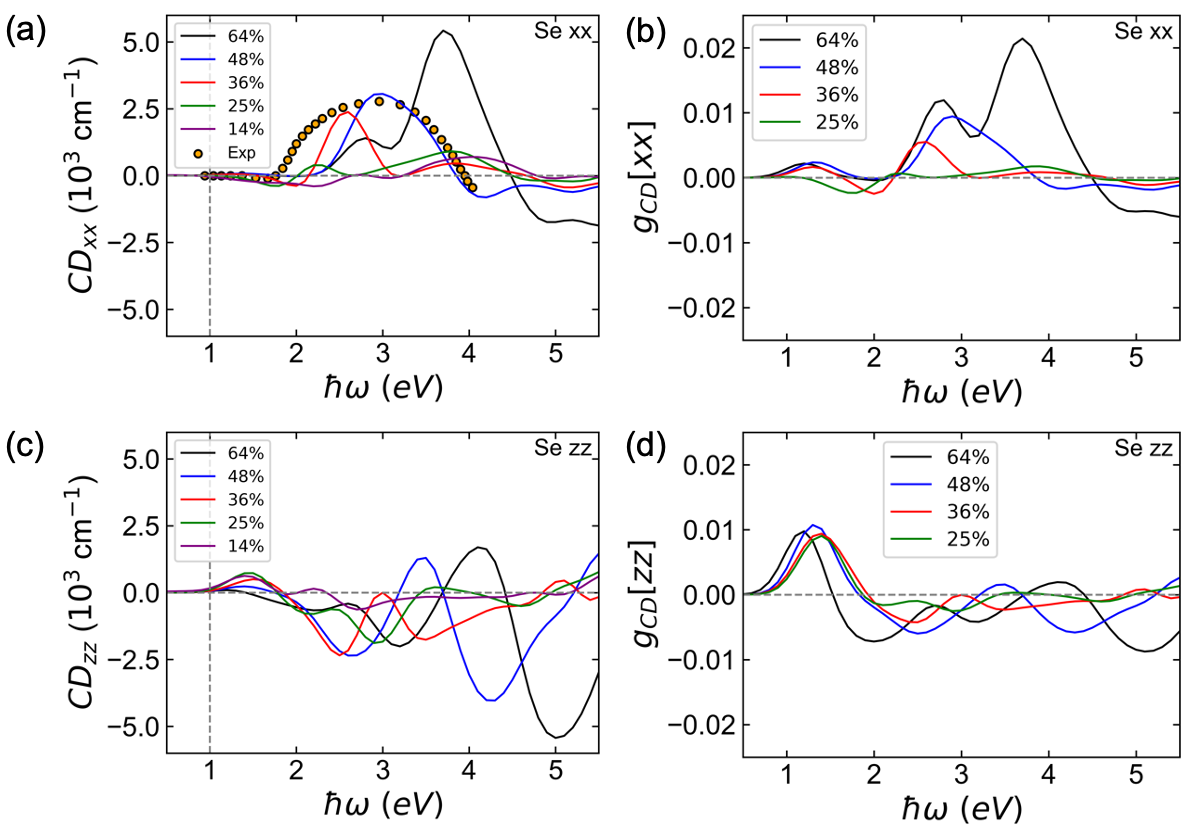}
    \caption{CD and $g_{CD}$ of the bulk Se system as a function of chirality. The experimental structure corresponds to chirality $48\%$. CD is reported in units of cm$^{-1}$,  see methods for details of unit definition. CD of bulk Se with different chiralities, (a) perpendicular to and (c) along the screw axis in z. (b) $g_{CD}$ of Se with different chiralities perpendicular to and (d) along the screw axis. Our calculated CD spectra (chirality $48\%$) agree with the experiment, denoted by the black circles in panel (a)\cite{saeva1977circular}. To enable clearer comparison, CD spectra were rigidly shifted along the energy and intensity axes for visualization purposes. The energy axes of all other chiralities were referenced to the spectrum of the configuration with \(u\)=0.218 ($S^2 = 48\%$).}

    \label{fig:Se-EQ-L-Sping-compare}
\end{figure}

Note that the total CD is defined as $CD=CD_{MD}+CD_{EQ}$. The $CD_{S}$ is currently included in $CD_{MD}$ unless specified. 
As shown previously,  contributions from the EQ are particularly important for anisotropic solids; their relation to chirality has yet to be investigated~\cite{multunas2023circular,Haque2024-yw}. We note that excitonic effects are not included in the current study, instead we focus on the above-bandgap transitions. As in our previous work~\cite{multunas2023circular,Haque2024-yw}, we find excellent agreement with experimental CD responses of above-bandgap transition in 2D hybrid perovskites and for the entire range in bulk Se, see Figures~\ref{fig:Exp-Benchmark}(d) and ~\ref{fig:Se-EQ-L-Sping-compare}(a).
We first confirm the systematic changes of CD with increasing chirality. We then provide an in-depth analyses of the dependence of chirality and various aspects of CD, including anisotropicity, MD/EQ contribution, spin/orbital components, and $g_{CD}$.

\begin{figure}
    \centering
    \includegraphics[scale=0.45]{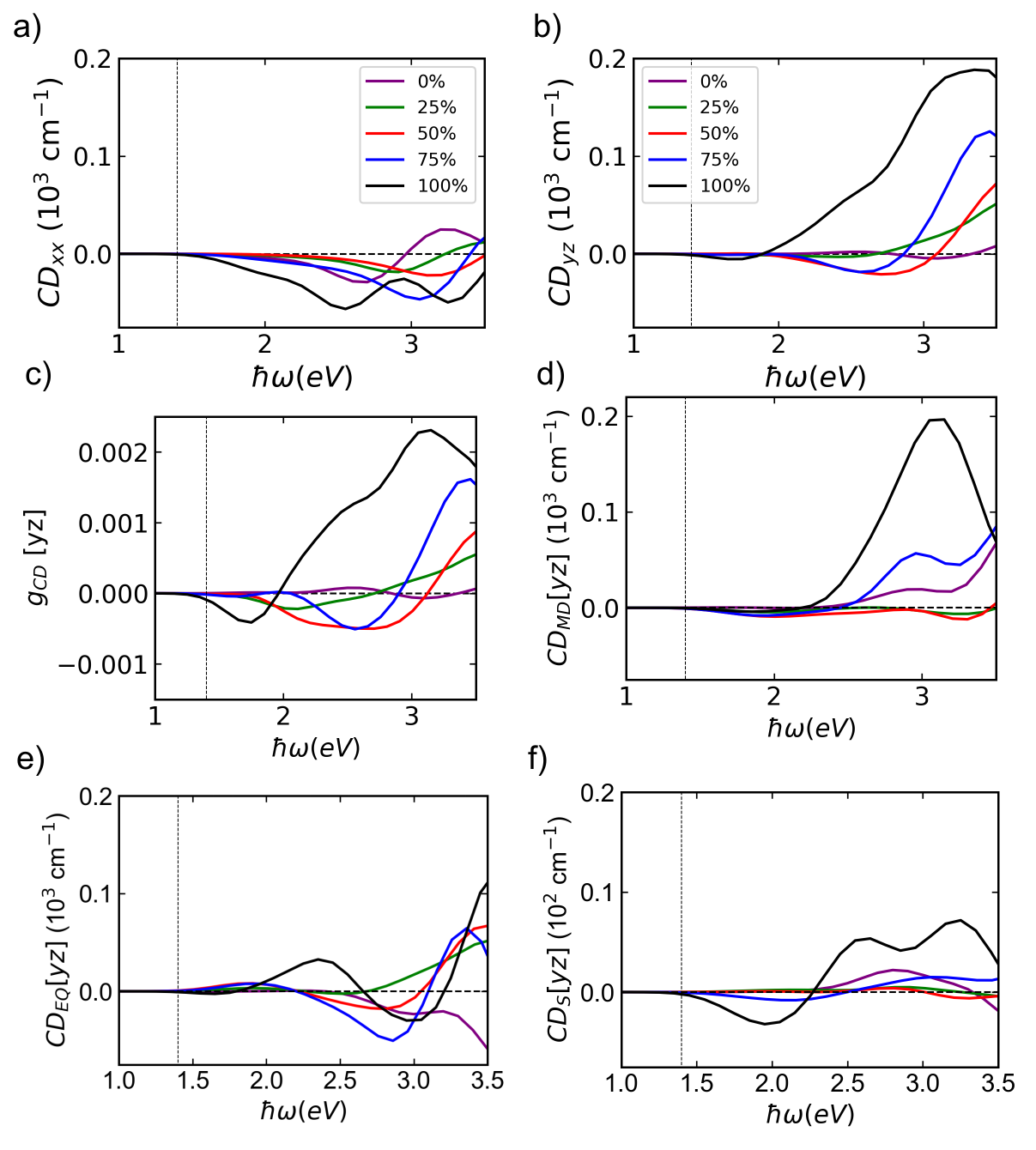}
    \caption{CD of the 2D-Cs-PB system as a function of chirality. CD is reported in units of cm$^{-1}$, see methods for details of unit definition. (a) CD response in the x-direction and (b) the yz component of the CD tensor. (c) $g_{CD}$ calculated from the yz component of the CD tensor. CD from different contributions, (d) magnetic diploe (MD), (e) electric quadrupole (EQ) and (f) spin with one order lower than the other components. Note that a rigid shift of spectra has been applied to align the bandgap with the 100\% chirality.}
    \label{fig:PVK-EQ-L-Sping-compare}
\end{figure}

In the 2D-Cs-PB system, the $P2_1$ screw axis is in-plane along the y direction formed by the equatorial Br ions (Figure~\ref{fig:Spin-texture-C-fitting}a), and leads to highly anisotropic CD responses as shown in SI Figure S8. The maximum response appears from the yz component of the CD tensor, and increases with the degree of chirality monotonically, while the other components of the CD tensor have a more subtle relation to chirality. This is because the chiral distortions are introduced not only into the equatorial Br but also to the axial Br; see SI Figure S6, which results in the yz component being most enhanced. Interestingly, in all other directions, minimal change in CD is observed until 100\% of the chiral distortion is introduced. Additionally, we can compute the $g_{CD}$ (defined as $g_{CD} = CD/\alpha$ where $\alpha$ is the optical absorption coefficient), which is an experimental technique that normalizes the CD by the optical absorption. The optical absorption coefficient is computed from the imaginary part of dielectric function: 

\begin{equation}
\begin{aligned}
\epsilon_{2_{\mu\nu}} =
\frac{4 \pi^2 e^2}{m_e^2 \omega^2}
\int& \frac{g_s \, d\mathbf{k}}{(2\pi)^3} 
\sum_{n',n}
\delta \!\left( \varepsilon_{\mathbf{k}n'} - \varepsilon_{\mathbf{k}n} - \hbar \omega \right)\\ 
&\times \left( f_{\mathbf{k}n} - f_{\mathbf{k}n'} \right)
(p^{\textbf{k}*}_{\mu,nn'}p^{\textbf{k}}_{\nu,nn'})~~,
\end{aligned}
\label{dielectric_function}
\end{equation}

where the variables' definition follows Eq.~\ref{CD_equation}. Then the optical absorption coefficient is defined based on the real part $\epsilon_{1}$ and the imaginary part $\epsilon_{2}$ of the dielectric function:

\begin{equation}
\alpha(\omega) \;=\;
\frac{\omega}{c \hbar} \cdot 
\frac{\varepsilon_2(\omega)}{
\sqrt{\dfrac{\varepsilon_1(\omega) + 
\sqrt{\varepsilon_1^2(\omega) + \varepsilon_2^2(\omega)}}{2}}}~~,
\end{equation}

where $c$ is the speed of light. 
The real part of the dielectric function ($\varepsilon_1$) is computed by applying the Kramers-Kronig relation to Eq.~\ref{dielectric_function}. In the 2D-Cs-PB systems CD and $g_{CD}$ have similar trends with respect to the degree of chirality, as shown in Figure~\ref{fig:PVK-EQ-L-Sping-compare}c. This is due to a nearly constant optical absorption with respect to chirality. 

To further examine the origin of CD increases with chirality, we decompose the contributions to CD into magnetic dipole, electric quadrupole, and spin contributions based on definitions in Eqs.~\ref{CD_MD_equation},~\ref{CD_EQ_equation},~\ref{CD_MD_equation2}, as shown in Figure~\ref{fig:PVK-EQ-L-Sping-compare}d, e, and f. Here we focus on the discussion on the yz component of the CD tensor given that it's the largest response,  although similar trends hold for other directions. 
The main contribution to CD is from the magnetic dipole ($CD_{MD}$ in Figure~\ref{fig:PVK-EQ-L-Sping-compare}d), with minor electric quadrupole contributions ($CD_{EQ}$ in Figure~\ref{fig:PVK-EQ-L-Sping-compare}e).  The spin contribution ($CD_{S}$ in Figure~\ref{fig:PVK-EQ-L-Sping-compare}f) is found to be one order of magnitude lower than the orbital contributions (MD+EQ).  
In the 2D-Cs-PB system, $CD$, $CD_{MD}$, $CD_{S}$, and $g_{CD}$ all increase with chirality, while no clear trend exists for the EQ contribution. The total CD is dominated by the MD contribution.

In bulk Se at the equilibrium structure, \(u\)=0.218 ($S^2 =48\%$), the computed CD spectrum agrees well with the experimental results in Ref.~\cite{saeva1977circular} (shown in Figure~\ref{fig:Se-EQ-L-Sping-compare}a). The original measurement was performed on the left-handed enantiomer along the x(y) direction, to compare with experiment, a path length of $l = 0.01\ \mathrm{mm}$ is assumed.
 
 In contrast to the 2D-Cs-PB system, bulk Se only has diagonal contributions to the CD tensor due to the \(D_3\) point group symmetry. The symmetry results in intrinsic anisotropy of the optical activity, with identical CD in the directions perpendicular to the screw axis (x and y) but different parallel to the screw axis (z),  (Figure~\ref{fig:Se-EQ-L-Sping-compare}a and c). The CD response exhibits a monotonic increase with chirality along the directions of both perpendicular and parallel to the screw axis. As in 2D-Cs-Pb, the results are dominated by the MD contribution, indicating that the MD contribution is more closely related to chirality than the EQ contribution across both systems.

In bulk Se, the $g_{CD}$ is found to have strong anisotropy, as in the CD response. However, only perpendicular to the screw axis (x), $g_{CD}$ exhibits a monotonic increase, and no trend is found parallel to the screw axis (z), as shown in Figure~\ref{fig:Se-EQ-L-Sping-compare}b and d. In fact, at low energies, $g_{CD}$ remains nearly unchanged along the screw axis as chirality increases. This is a key result showing that $g_{CD}$ is not always a good measure of the degree of chirality. Often $g_{CD}$ is viewed as a better metric than CD for quantifying chirality because $g_{CD}$ is normalized by total absorption. The optical absorption is largely determined by the first-order (electric dipole) term in the dielectric function. For CD of non-magnetic systems, this term is zero due to time-reversal symmetry; see~\cite{multunas2023circular} for details. This leads to CD and total absorption having a different dependence on the electric dipole. Depending on how the electric dipole changes with chirality (there is no direct connection between the two in principle), this can lead $g_{CD}$ to increase, decrease, or be invariant with respect to chirality, as seen in bulk Se. 
Although $g_{CD}$ is useful for comparing the chiral-optical responses of different materials~\cite{song2024enhancing}, it cannot be used as a universal quantitative measure of the degree of chirality in general materials.

Additionally, we note that such continuous tuning of chirality in Se can be realized by applying strain \cite{fecher2022chirality, keller1977effect} as shown in SI Figure S9. As the pressure increases from 0 to 1 and then to 4 GPa, the corresponding chiralities decrease from 48\%, to 35\%, and finally to 22\%, which is determined by the $3a$ Wycoff position parameter \(u\) for $P3_121$ symmetry. The variation of CD and $g_{CD}$ under pressure as a function of chirality in SI Figure S9 is similar to the unstressed case discussed earlier.

\subsection{Effect of Chirality on CPGE}

\begin{figure*}[!hbt]
    \centering
    \includegraphics[scale=0.55]{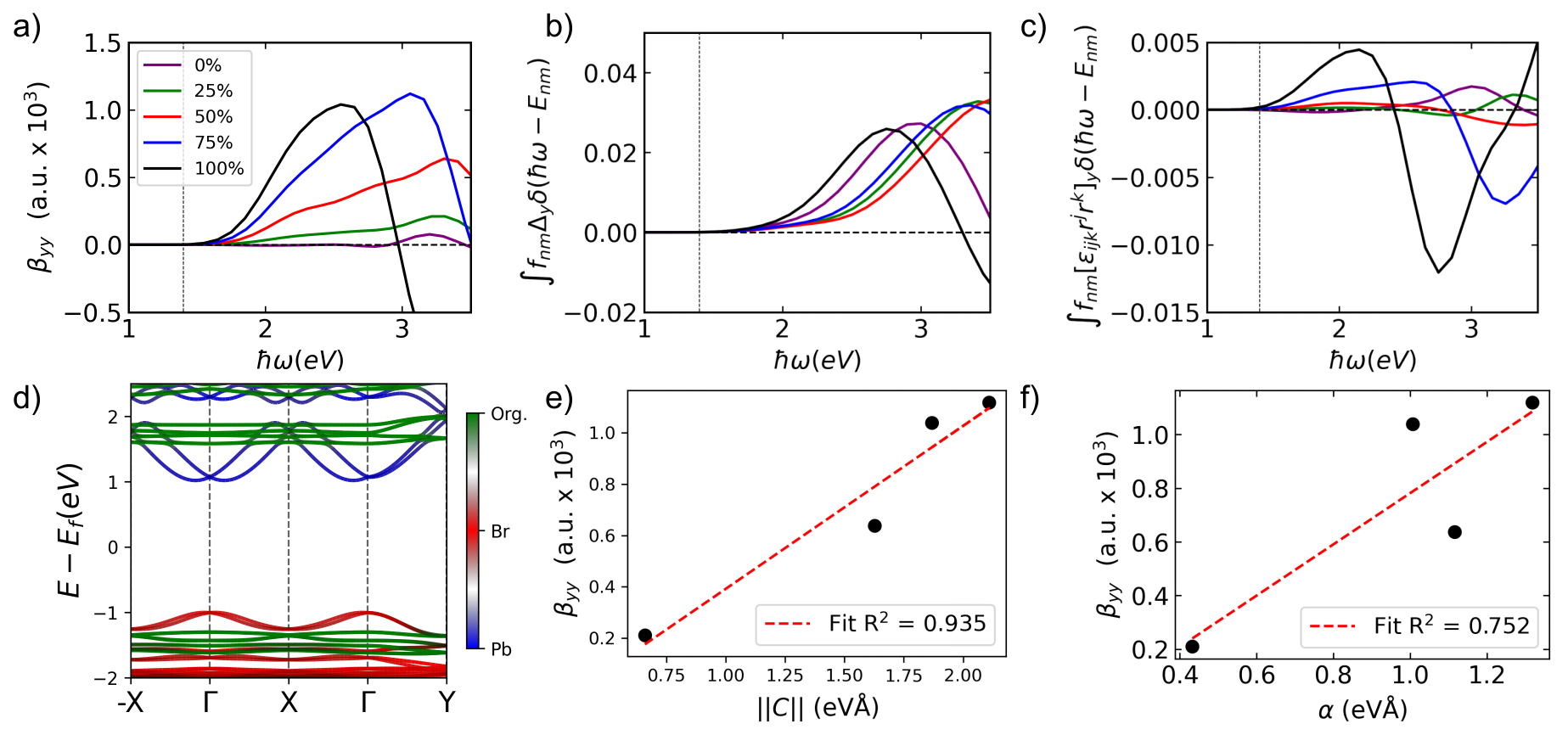}
    \caption{CPGE analysis as a function of chirality for 2D-Cs-PB. (a) CPGE tensor as a function of chirality. Contribution from (b) electron-hole group velocity difference and (c) dipole-matrix elements. Note in panels (a-c), a rigid shift has been applied to the excitation energies to align the bandgap with the 100\% chiral structure. (d) Bandstructure of 2D-NPB with color indicating contributions from Pb, Br, and organic (Org.). Correlation of CPGE to (e) the total SOC computed form the norm of the $C$ tensor and (f) $\alpha$ coefficient fit using the Rashba Hamiltonian in Eq.~\ref{RashbaH}. The linear regression of calculated points in (e) and (f) is plotted as a red dashed line.}
    \label{fig:Se-PVK-CPGE-parts}
\end{figure*}

Circularly polarized light-induced DC photocurrent is present in certain non-centrosymmetric materials, and known as the circular photogalvanic effect, CPGE. 
Although chirality is not required for CPGE,  chiral structures will exhibit CPGE since they lack both inversion and mirror symmetries.
Experimentally, CPGE is often used to probe the amount of spin splitting, the type of spin-orbit coupling~\cite{yu2015temperature,zhu2024electrically}, or the nontrivial topology~\cite{de2017quantized,le2020ab}. 
We discuss the physical mechanism and quantitative relationship between interband CPGE and the degree of chirality CCM, in terms of the magnitude and symmetry of interband CPGE, the contributions from individual components, and its relation to SOC.

Near the band edge, only transitions from the VBM to the CBM are allowed, below approximately 2.5 eV.
Since the split of these pairs of bands are intuitively described by SOC, we relate the band edge CPGE responses to the spin splitting and SOC. We focus on  2D-Cs-PB here and analysis of Se can be found in SI section V. The maximum of the first peak in Figure~\ref{fig:Se-PVK-CPGE-parts}a is plotted against the SOC properties in Figure~\ref{fig:Se-PVK-CPGE-parts}e and f. This maximum corresponds to the highest energy where only VBM to CBM transitions occur. (Note this is slightly different at different chiralities, see Figure S2). There is a general correlation with SOC and CPGE at the band edge. We compare correlation with chirality to the extracted total SOC (norm of the full $C$ tensor as described in Section 2.2,) and the $\alpha$ parameter defined by the eigenvalues of the Rashba Hamiltonian:

\begin{equation}
    E^{\pm}(k_x) = \frac{\hbar^2k_x^2}{2m^*}\pm \alpha k_x \label{RashbaH}
\end{equation}

where $m^*$ is the carrier effective mass and $E^{\pm}$ is the energies of the spin-split bands. For both methods the general trend is similar, namely CPGE increases with increasing Rashba splitting (see SI for more details). We find total SOC has significantly better correlation with CPGE (R$^2$ = 0.935), compared to the $\alpha$ parameter (R$^2$ = 0.752) defined in Eq.~\ref{RashbaH}. 
Note that computing the $\alpha$ parameter from Eq.~\ref{RashbaH} assumes that the spin splitting is entirely from Rashba-type SOC, which does not hold for this system as shown in Figure~\ref{fig:Spin-texture-C-fitting}d. This indicates that CPGE magnitude near the band edge is closely related to the \emph{total SOC}, not just the Rashba parameter $\alpha$. 
  
Given that in this work CPGE is computed by the Kubo formula with the relaxation time approximation in Eq.~\ref{eq:cpge}, both light absorption (electric transition dipoles $r^k_{\boldsymbol{k},nm}$) and electron-hole group velocity difference ($\Delta^i_{\boldsymbol{k},nm}$) determine the magnitude of CPGE. Again focusing on the band edge transitions, as chirality increases there is an increase in the spin splitting as well as changes to the band dispersion.  This results in a minimal change in $\Delta^i_{\boldsymbol{k},nm}$ and no clear correlation with chirality 
(e.g. even with no chirality present there is still a substantial $\Delta^i_{\boldsymbol{k},nm}$). 
Meanwhile, due to the increased spin splitting,   electric transition dipoles near the band edge show a systematic increase with chirality as seen in Figure~\ref{fig:Se-PVK-CPGE-parts}c. This corresponds to a higher transition probability with respect to circularly polarized light, and is responsible for the increase in CPGE with increasing chirailty for excitation energies less than 2.5 eV.

For transitions well above the band edges,  the CPGE includes transitions between many bands and can no longer easily be correlated with spin splittings of band edges or particular bands. When other bands begin to contribute at high excitation energies, significant electron-hole group velocity differences begin to emerge, although without a clear trend with chirality, as seen above 3 eV in Figure~\ref{fig:Se-PVK-CPGE-parts}b. The dipole contribution tends to increase with increasing excitation energies due to a larger joint density of state, which is consistent with the increase in magnitude above 2.5 eV excitation energy in Figure~\ref{fig:Se-PVK-CPGE-parts}c. 
The combination of these two trends at higher excitation energies ultimately removes a correlation between chirality and CPGE. This highlights the complicated interplay between the individual components of CPGE and the overall trend with chirality. We conclude that CPGE is more strongly correlated with total SOC than chirality, in particular close to band edge transitions.yz

\subsection{Chirality Transfer through Electronic and Structural Effects on Optical Activity}

\begin{figure}
    \centering
    \includegraphics[scale=0.4]{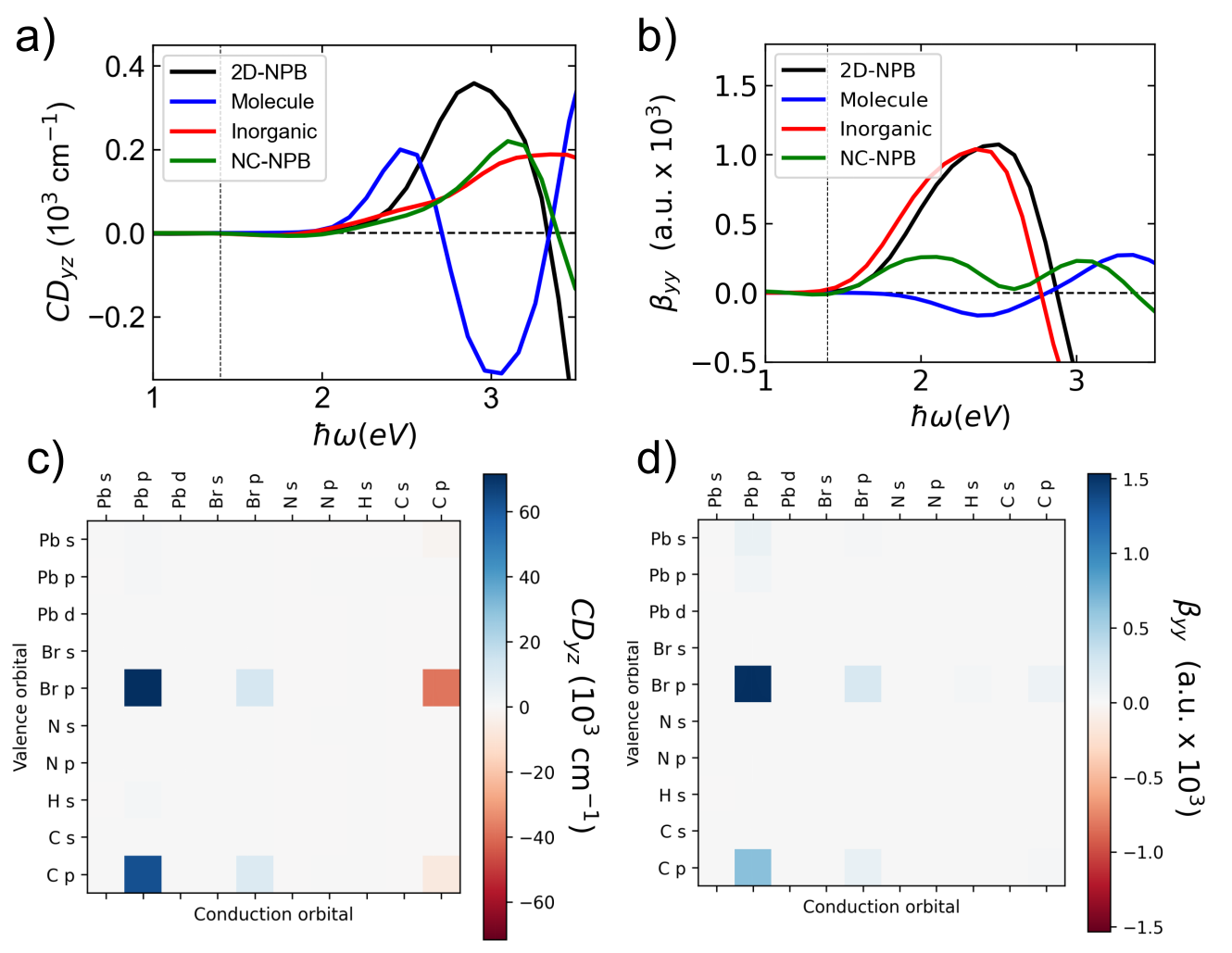}
    \caption{CD and CPGE in 2D-NPB and their transition analysis. CD is reported in units of cm$^{-1}$, see methods for details of unit definition.(a) CD and (b) CPGE in 2D-NPB for with separated contributions from the inorganic, molecule, and with the molecules present but no chirality in the inorganic layer (NC-NPB). Note all energies have been shifted to the minimum bandgap indicated by the dashed line. Transition analysis (c) for CD at 3.0 eV and (d) for CPGE at 2.4 eV in 2D-NPB.}
    \label{fig:CD-CPGE-parts}
\end{figure}

The CD response in 2D chiral perovskites can have contributions from both the chiral distorted inorganic lattice and chiral molecules themselves. To decompose the relative contributions we examine four systems: the molecule system alone (Molecule), the chiral perovskite sublattice (Inorganic), the combined molecule and inorganic sublattice (2D-NPB) and the combined system where we restore the original perovskite symmetry by removing chirality from the inorganic layer, restoring perfect cubic symmetry (NC-NPB).  Looking at the separate contributions, the molecular component (Molecule) exhibits a CD response much larger than that of the inorganic component alone (Inorganic). However, when the molecular and inorganic parts are combined into a hybrid system, the overall CD signal is notably enhanced. The amplification of CD suggests a synergistic interplay between the structural chirality and the electronic contributions. This is further supported by the reduction in CD when the chiral distortions are removed from the inorganic sublattice (NC-NPB), Figure~\ref{fig:CD-CPGE-parts}a. The enhancement of CD from the combination of structural distortion and electronic effects is consistent with previous work on remote chirality transfer from a chiral molecule to achiral perovskites~\cite{Haque2024-yw}.
Specifically, the electronic interactions, such as charge transfer or orbital overlap, along with the structural distortion introduced by molecule-inorganic lattice interactions, determine the overall CD response. These factors are critical to understanding the signatures of CD in the hybrid system. 

We further analyze CD at 3.0 eV by projecting the pair of states in the transition matrix elements to atomic orbitals, as shown in Figure~\ref{fig:CD-CPGE-parts}c. The analysis reveals three dominant transitions that correspond to the main peak at 3.0 eV. The strongest contribution comes from the transition from Br-$p$ to Pb-$p$ orbitals, followed in intensity by C-$p$ to Pb-$p$ and Br $p$ to C-$p$. Interestingly, the sign of the Br-$p$ to C-$p$ transition is opposite to the other main contributions and results in CD reduction. However, the most dominant contribution comes from the states of the inorganic lattice. Such an analysis highlights the enhancement of the CD response in inorganic lattices that require wavefunction overlaps with organic molecules.    

In contrast to CD, the CPGE is dominated mainly by contributions from the inorganic sublattice. As seen in Figure~\ref{fig:CD-CPGE-parts}b, the inorganic contribution almost reproduces exactly the combined organic-inorganic response. This can be explained by the composition of the band-edge states which are mostly inorganic in character.  Another noticeable difference between CD and CPGE is the onset of CPGE started at much lower excitation energies i.e. 1.5 eV compared to CD (its onset above 2 eV). 

Similar to CD analysis, we project the transition matrix elements onto the specific transitions and find the majority of the CPGE is from the inorganic (Br-$p$ to Pb-$p$), with minor contributions from C-$p$ to Pb-$p$, as shown in Figure~\ref{fig:CD-CPGE-parts}d. More importantly, when removing the chiral distortion from the inorganic sublattice but keeping organic molecules present, we see almost no CPGE. This indicates that the chiral distortion to the inorganic sublattice is critical for controlling CPGE in 2D-hybrid perovskites.

\section{Conclusion}
In this work we reveal the correlation between the continuous-chirality measure (CCM) with spin, orbital polarization, and optical activity of solids from first-principle calculations. First, through quantitative analysis of spin textures, we find that the spin-orbit field or effective internal field monotonically increases with increasing chirality. The specific SOC components vary rather differently among the two systems. Meanwhile, we find that orbital angular momentum textures closely resemble those of spins.
The optical activity such as circular dichorism is confirmed as a direct measure of chirality, across different types of chiral materials; however, we emphasize the anisotropy of CD needs careful consideration. Specifically, through examination of different components of CD, we find MD contributions show correlation with chirality, but not EQ and spin. Another key result is that the absorption dissymmetry factor, $g_{CD}$, is not always a good chirality metric. This needs to be considered when using $g_{CD}$ as a direct measure of chirality experimentally.
On the other hand, the CPGE does not show a consistent correlation with chirality, except at the low energy transition close to band edges. Even though spin splitting may be enlarged with increasing chirality, which correlates with CPGE responses close to the band edge, other factors, in particular dominant at high excitation energies, such as electric transition dipoles or electron-hole velocity difference may have the opposite trends. These results highlight the complex interplay of chirality with $g_{CD}$ and CPGE. Overall we find that CD has the strongest correlation with chirality.   

In addition to the role of chirality, we investigate the contributions to CD and CPGE from different components of the 2D-HOIP, which provides further insight into the role of the organic and inorganic components since they both have chirality.  
We demonstrate the significant role of synergistic relationship between structural chirality and electronic effect through organic-inorganic interactions in producing the maximum CD. In contrast, the chirality transfer to the inorganic sublattice is more dominant for CPGE, with little direct contribution from molecules. 
This work provides critical insights into the role of chirailty on spin, orbital textures, chiral-optical properties, which can be guidelines of controlling these properties through strain engineering or chirality transfer at interfaces.  

\section{Computational Methods}
The electronic structure and geometry optimization calculations were performed in the open-source plane-wave DFT software JDFTx~\cite{sundararaman2017jdftx}. We employed norm-conserving pseudopotentials from PseudoDojo~\cite{van2018pseudodojo}, using scalar-relativistic and fully relativistic treatments for calculations without and with spin-orbit coupling, respectively, with the plane wave basis set defined by a kinetic energy cutoff of 60 Ha for Se and 50 Ha for the 2D-perovskites. The exchange-correlation functional used in all calculations was the Perdew-Burke-Ernzerhof (PBE) generalized gradient approximation~\cite{perdew1996generalized}. The initial structure of hybrid perovskites is from Ref.~\cite{jana2020organic}, and Se is from the Materials Project~\cite{jain2013commentary}. See Supplemental Material for more structural information and additional computational details.
For Brillouin zone sampling in self-consistent calculations, we used a \(\Gamma\)-centered k-mesh. Specifically, a \(5 \times 5 \times 1\) mesh was used for 2D hybrid perovskites, while a \(12 \times 12 \times 12\) mesh was used for Se. Self-consistent calculations were followed by non-self-consistent calculations with high density k-points for the evaluation of matrix elements. 

We employed Monte-Carlo sampling within the first Brillouin zone. In a crystalline system, \( L \), \( Q_{\mu\nu} \) and \( R \) elements  often exhibit sharp peaks rather than varying smoothly with \(k\), thus special treatment is required to ensure computational accuracy~\cite{multunas2023circular}.

\begin{figure}
    \centering
    \includegraphics[scale=0.42]{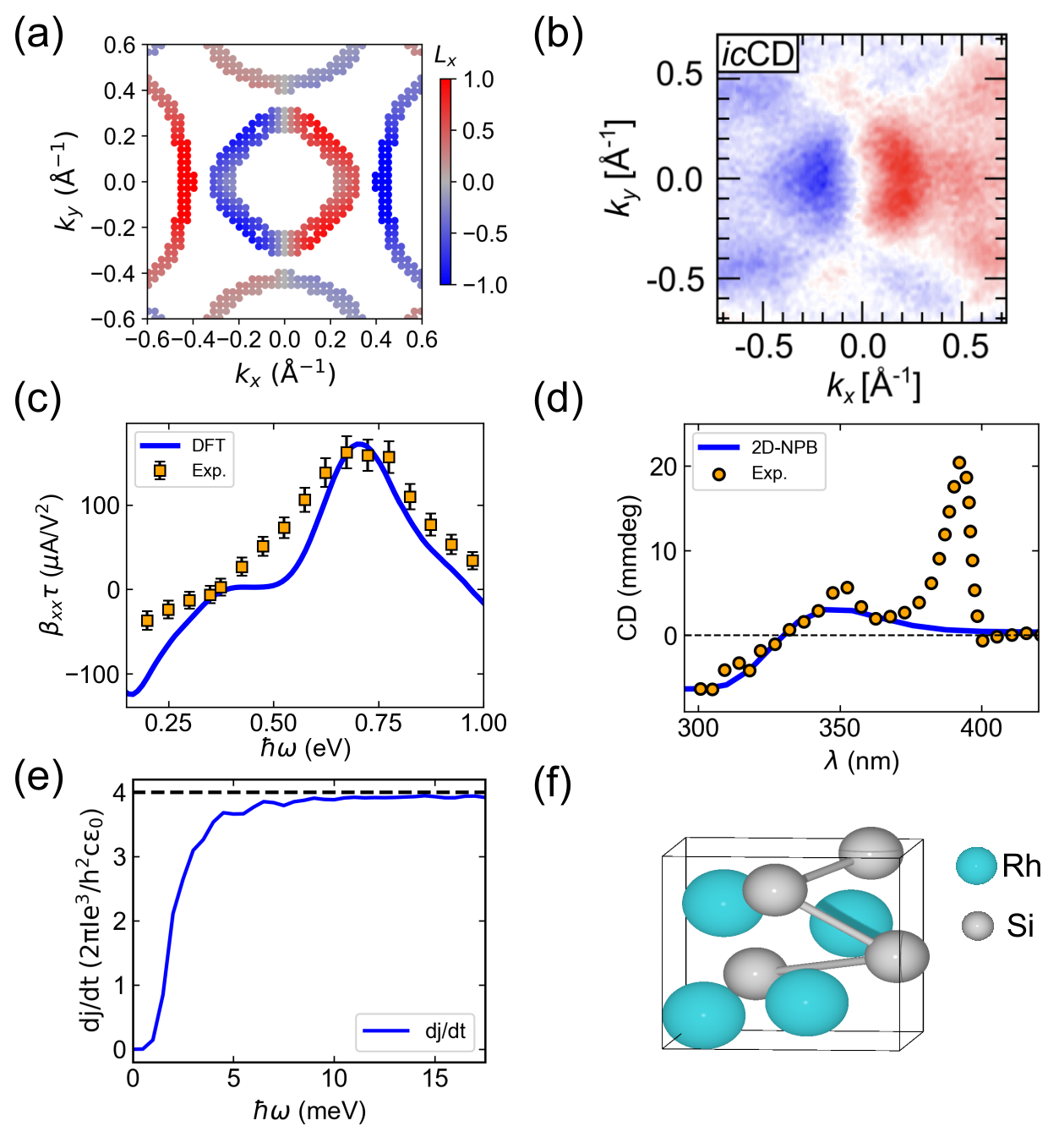}
    \caption{Comparison between first-principles calculations and experimental results on OAM, CD, and CPGE. (a)$\mathrm{L}_x$ at 0.25 eV below the Fermi level in CoSi calculated from first-principles and (b) measured by CD-ARPES. The CD-ARPES results were adapted with permission from Ref.~\cite{brinkman2024chirality}. (c) CPGE in RhSi calculated using a relaxation time 6.6 fs is compared with experimental measurements~\cite{ni2020linear}. (d) Comparison of calculated CD of 2D-NPB with experiments~\cite{jana2020organic}. Since the experimental sample is polycrystalline, we compute the isotropic average of the CD tensor and convert from $\Delta \alpha$ to mmdeg, with a path length of 0.02 mm. Additionally, a scissor of 0.6 eV (rigid shift of the conduction band) was used to match the experimental bandgap of 2D-NPB. (e) Quantized CPGE of RhSi at low excitation energies. A quantization of 4 is obtained from first-principles calculations. (f) RhSi atomic structure.}
    \label{fig:Exp-Benchmark}
\end{figure}

To benchmark the implementation of OAM based on Eq.(5), we compare OAM in the chiral topological semimetal CoSi between theory and experiments. The calculated OAM distribution in the $\Gamma$–$\text{X}$–$\text{M}$ plane at 0.25 eV below the Fermi level, as shown in Figure~\ref{fig:Exp-Benchmark}a, matches well with CD-ARPES experiments in Figure~\ref{fig:Exp-Benchmark}b~\cite{brinkman2024chirality}. This validates our \( L \) matrix implementation.

The circular dichroism (CD) is calculated as the differential Naperian absorbance $\Delta \alpha$, with units of {cm$^{-1}$}, which quantifies the difference in absorption between left- and right-circularly polarized light. In experiments on crystals, CD is typically reported as the absolute ellipticity $\theta$ (\textit{e.g.}, in degrees), which is often given without specifying the corresponding path length $l$. To compare with experiments, it is essential to relate the computed $\Delta \alpha$ to the measured $\theta$. When expressed per unit path length $l$, the conversion is:
\begin{equation}
\frac{\theta}{l} \left[ \frac{\mathrm{deg}}{\mathrm{mm}} \right] = \frac{45}{10\pi} \cdot \Delta \alpha \left[ \mathrm{cm}^{-1} \right] \approx 1.4324 \cdot \Delta \alpha
\end{equation}
Additional theoretical details on the CD implementation can be found in Ref.~\cite{multunas2023circular}.

The CPGE tensor was implemented in an in-house code FP-CPGE (https://github.com/Ping-Group-UCSC/FP-CPGE) that is interfaced with JDFTx. CPGE in this work was computed for circularly-polarized light-induced injection current, derived from the second-order perturbation theory under the relaxation-time approximation:

\begin{equation} 
\label{eq:cpge}
\beta_{ij} =A \epsilon_{jkl}\sum_{\boldsymbol{k},n,m}f^{\boldsymbol{k}}_{nm}\Delta^i_{\boldsymbol{k},nm}r^k_{\boldsymbol{k},nm}r^l_{\boldsymbol{k},mn}\delta(\hbar\omega-E_{\boldsymbol{k},nm})
\end{equation}
\begin{equation}
A=\frac{e^3 \pi}{N_{\boldsymbol{k}}V\hbar}
\end{equation}

where $n$ and $m$ are band indices and $i,j,k$ and $l$ are Cartesian directions. Here, $f_{nm}$ is the occupation difference,  $E_{\boldsymbol{k},nm}$ is the energy difference, $r^k_{\boldsymbol{k},nm}$ are the dipole-matrix elements, and $\Delta^i_{\boldsymbol{k},nm}$ are the difference in velocity-matrix elements of band $n$ and $m$ at k-point $\boldsymbol{k}$. All these elements are computed from DFT directly, with Monte-Carlo k-point samplings, without Wannier interpolations. Convergence with respect to k-point sampling was achieved at 2,000 k-points, see SI Figure S11c and d.

We benchmark the CPGE implementation against experiments for RhSi ~\cite{ni2020linear} in Figure~\ref{fig:Exp-Benchmark}c and e. The experiment estimates a relaxation time $\tau$ of 6.6 fs which is used in our calculations. We obtain good agreement with experiment with respect to the maximum peak position at 0.7 eV and magnitude of 150 $\mu AV^{-2}$, as seen in Figure~\ref{fig:Exp-Benchmark}c. Additionally, we obtain excellent agreement with the zero crossing around 0.4 eV. As an additional benchmark, we compute the low energy CPGE in RhSi in Figure~\ref{fig:Exp-Benchmark}e. 
It has been proposed RhSi owns quantized CPGE at the low energy range due to its topological properties 
~\cite{de2017quantized}. When the lower-energy crossing is fully occupied and cannot be excited, the CPGE is proportional to the Chern number of the accessible band crossing. 
In previous work, a tight-binding model was employed for the CPGE of RhSi, obtaining a quantization at 4.  
This is consistent with the Chern number of the band-crossings near the $\Gamma$ high-symmetry point ~\cite{chang2017unconventional}.  We obtain a quantization at 4 from excitation energy 8-17 meV as shown in Figure~\ref{fig:Exp-Benchmark}e.

Finally we compare the computed CD for 2D-NPB with experiments. We find excellent agreement with the features above the bandgap transitions. Given that we did not include many-body effects such as electron-hole exchange interaction in the current CD calculations, the excitonic peak below 400nm in experiments is not expected to reproduce. Since the experimental sample is polycrystalline, we compute the isotropic average of the CD tensor, and to convert from $\Delta \alpha$ to mmdeg, with a path length of 0.02 mm. Additionally a scissor of 0.6 eV to rigid shift conduction band was used, in order to match the experimental bandgaps. More benchmarks of CD against experiments can be found in Ref.~\cite{multunas2023circular}, which will not be repeated here.

\section*{acknowledgement}
We acknowledge support for the theoretical and code development of optical activity, CPGE, and CCM by the computational chemical science program within the Office of Science at DOE under grant No. DE-SC0023301. 
We acknowledge the support for materials theory application as part of the Center for Hybrid Organic-Inorganic Semiconductors for Energy (CHOISE), an Energy Frontier Research Center funded by the Office of Basic Energy Sciences, Office of Science within the US Department of Energy (DOE). 
This research used resources of the Scientific Data and Computing center, a component of the 
Computational Science Initiative, at Brookhaven National Laboratory under Contract No. DE-SC0012704,
the National Energy Research Scientific Computing Center (NERSC) a U.S. Department of Energy Office of Science User Facility operated under Contract No. DE-AC02-05CH11231. 
This work used the TACC Stampede3 system at the University of Texas at Austin through allocation PHY240212 from the Advanced Cyberinfrastructure Coordination Ecosystem: Services and Support (ACCESS) program~\cite{Boerner2023}, which is supported by US National Science Foundation grants No. 2138259, No. 2138286, No. 2138307, No. 2137603, and No. 2138296.

\bibliographystyle{apsrev4-2}
\bibliography{ref}
\end{document}